\documentstyle[aps,epsf,rotate]{revtex}
\begin{document}
\bibliographystyle{prsty}
\draft
\renewcommand{\thefootnote}{\fnsymbol{footnote}}
\title{The Nos\'e--Hoover thermostated Lorentz gas
\footnote{Note: Most of the figures are in poor quality output. The originals are many MB large. They can be obtained upon request.}}
\author{K. Rateitschak and R. Klages}
\address{Center for Nonlinear Phenomena and Complex Systems, 
Universit\'{e} Libre de Bruxelles,\\ 
Campus Plaine CP 231, Blvd du Triomphe, B-1050 Brussels, Belgium,
e-mail:~krateits@ulb.ac.be}  
\author{W. G. Hoover}
\address{Department of Applied Science, University of California at
Davis/Livermore,\\Livermore, CA 94551-7808, USA}
\date{\today}
\maketitle

\begin{abstract}
We apply the Nos\'e--Hoover thermostat and three variations of it,
which control different combinations of velocity moments, to the
periodic Lorentz gas. Switching on an external electric field leads to
nonequilibrium steady states for the four models with a constant
average kinetic energy of the moving particle.  We study the
probability density, the conductivity and the attractor in
nonequilibrium and compare the results to the Gaussian thermostated
Lorentz gas and to the Lorentz gas as thermostated by deterministic
scattering.
\end{abstract}

\section{introduction}
In a system of particles under an external force a nonequilibrium
steady state can be obtained by applying a thermostat
\cite{EvMo90,Hoov91,MoDe98}. Deterministic and time reversible bulk
thermostating is based on introducing a momentum dependent friction
coefficient in the equations of motion.  One type of this mechanism is
the Nos\'e--Hoover thermostat \cite{Nose84b,Hoov85}. It creates a
canonical ensemble in equilibrium and yields a stationary
nonequilibrium distribution of velocities in nonequilibrium.  Another
version, the Gaussian isokinetic thermostat \cite{HLM82,Ev83,EH83},
leads to a microcanonical density for the velocity components in
equilibrium and to a constant kinetic energy in nonequilibrium.
Though the microscopic dynamics of these thermostated systems is time
reversible the macroscopic dynamics is irreversible in nonequilibrium
\cite{HHP87}. This is related to a contraction onto a fractal
attractor \cite{MH87,Morr89,HoB99}.

Characteristic features of thermostated many particle systems, like a
nonequilibrium steady state and a fractal attractor, have been
recovered for a specific one particle system, the Gaussian
thermostated Lorentz gas
\cite{MH87,LRM94,LNRM95,DeGP95,DettmannMorriss96}.  The periodic
Lorentz gas consists of a particle that moves through a triangular
lattice of hard disks and is elastically reflected at a collision with
a disk\footnote{A model almost identical to the driven periodic
Lorentz gas, except for some geometric restrictions, is the Galton
board, which has been invented in 1873 to study probability
distributions \cite{kac64}.}. It serves as a standard model in the
field of chaos and transport, see e.~g., \cite{MoDe98,MH87}.  In
contrast to many particle systems a one particle system reflects more
strongly the properties of a thermostat. For the Gaussian thermostated
Lorentz gas a complicated dependence of the attractor on the field
strength results \cite{MH87,LRM94,LNRM95}.

A second simple one particle system which has been much investigated
concerning its chaotic properties is the Nos\'e--Hoover thermostated
harmonic oscillator. In contrast to the Lorentz gas, the dynamics of
this system is generically nonergodic
\cite{PHV86,PoschHoover97}. However, one can obtain an ergodic
dynamics for this system by the additional control of the square of
the kinetic energy \cite{PoschHoover97,HoHo96,HoKu97}.
 
Recently an alternative thermostating mechanism, called thermostating
by deterministic scattering, has been introduced for the periodic
Lorentz gas \cite{KRN98,RKN99}\footnote{This mechanism has been
applied later to a system of hard disk under a temperature gradient
and shear\cite{WKN98}.}. This deterministic and time reversible
mechanism is based on including energy transfer between moving
particle and disk scatterer at a collision, instead of using a
momentum dependent friction coefficient. It leads to a canonical
probability density for the particle in equilibrium, and in
nonequilibrium it keeps the energy of the particle on average
constant. In Refs.~\cite{KRN98,RKN99} this model has been compared to
the Gaussian thermostated Lorentz gas: In nonequilibrium one finds an
attractor for this model which is similar to the fractal attractor of
the Gaussian thermostated Lorentz gas, but in contrast to the Gaussian
case the attractor is phase space filling even for high field
strengths. For both models the conductivity is a nonlinear decreasing
function with increasing field strength on a coarse scale.

Based on its construction the method of thermostating by deterministic
scattering is in fact closer to the Nos\'e--Hoover thermostat.  This
motivates us to apply for the first time the Nos\'e--Hoover thermostat
to the periodic Lorentz gas.  In Section II we introduce the
Nos\'e--Hoover thermostat and discuss some variations of it.  In
Section III we define the periodic Lorentz gas and the thermostats we
will study.  We investigate these models in equilibrium in section IV
and in nonequilibrium in Section V. Conclusions are drawn in Section
VI.

\section{The Nos\'e--Hoover thermostat and some variations}
In the following sections we consider a one particle system in two
dimensions with the position coordinates $\vec{q}=(q_x,q_y)$ and the
momentum coordinates $\vec{p}=(p_x,p_y)$. The mass of the particle has
been set equal to unity.  The equations of motion for the
Nos\'e--Hoover thermostat are then given by \cite{Hoov85}
\begin{eqnarray}
\dot{\vec{q}} &=& \vec{p} \nonumber\\
\dot{\vec{p}} &=& \vec{\varepsilon}-\zeta\vec{p}\nonumber\\
\dot{\zeta}&=&(\frac{p^2}{2T}-1)\frac{1}{\tau^2}\;.
\label{nh}
\end{eqnarray}
The thermostat variable $\zeta$ couples the particle dynamics to a
reservoir. It controls the kinetic energy of the particle $p^2/2$ such
that $<p^2>=2T$.  This holds even in nonequilibrium as induced by an
electric field $\vec{\varepsilon}$.  $\tau$ is the response time of
the thermostat. Performing the limit $\tau\to 0$ in Eqs.~(\ref{nh})
approximates the Gaussian thermostat with
$\zeta=(\vec{\varepsilon}\cdot\vec{p})/p^2$.  In the limit
$\tau\to\infty$ the friction coefficient approaches a constant,
$\zeta=\zeta_c$, and the equations of motion are not time reversible
anymore. The dynamics of this dissipative limit has been investigated
in \cite{HoMo92}.

A generalization of the Nos\'e--Hoover thermostat to control higher
even moments of $p$ has been introduced by Hoover \cite{Ho89}. The
moments are fixed according to the momentum relations for the Gaussian
distribution. By such a more detailed control of the nonequilibrium
steady state statistical dynamical properties, like ergodicity, can be
improved, as has been mentioned for the Nos\'e--Hoover thermostated
harmonic oscillator in the introduction.  In principle the method of
the control of the even moments can straightforwardly be extended to
the control of the odd moments.  However, such a thermostat would
involve an additional parameter to control the respective current of
the subsystem.  Thus, the corresponding reservoir would be more than a
single thermal reservoir, which is physically not desirable, apart
from the fact that the respective equations of motion would not be
time reversible anymore.

We briefly note that there exist other formal generalizations
\cite{JeBe88}, or modifications \cite{BuKu90,MKT92}, of the Nos\'e--Hoover
thermostat in the literature. They have been critically reviewed in
Refs.~\cite{HoHo96,HoKu97,Ho89}.

\section{Variations of the Nos\'e--Hoover thermostat for the  periodic
Lorentz gas}

The basic themostating method we investigate in this paper is the
Nos\'e--Hoover thermostat, Eqs.~(\ref{nh}).  In the following we
introduce three variations of it which are time reversible and result
in a dissipative dynamics in nonequilibrium.
 
The {\em first variation} is the Nos\'e--Hoover thermostat with a
field dependent coupling to the reservoir,
\begin{eqnarray}
\dot{\vec{q}} &=& \vec{p} \nonumber\\
\dot{p_x} &=& \varepsilon_x-(1+\varepsilon_x)\zeta p_x \nonumber\\
\dot{p_y} &=& \varepsilon_y-(1+\varepsilon_y)\zeta p_y \nonumber\\
\dot{\zeta}&=&(\frac{p^2}{2T}-1)\frac{1}{\tau^2}\;,
\label{nhe}
\end{eqnarray}
which is obtained by including the factors $1+\varepsilon_x$,
resp.~$1+\varepsilon_y$, in Eqs.~(\ref{nh}).  Alternatively, these
equations can be written by defining two field dependent friction
coefficients, $\xi_x=(1+\varepsilon_x)\zeta$ and
$\xi_y=(1+\varepsilon_y)\zeta$, which are governed by
$\dot{\xi}_x=(p^2/2T-1)(1+\varepsilon_x)/\tau^2$ and $\dot{\xi}_y
=(p^2/2T-1)(1+\varepsilon_y)/\tau^2$, respectively. It then becomes
clear that for each momentum component there exists a separate
response time of the reservoir which reads
$\tau/\sqrt{1+\varepsilon_x}$,
resp.~$\tau/\sqrt{1+\varepsilon_y}$. The basic advantage of this
thermostat is that the response times are now adjusted to the
corresponding component of the field strength such that with
increasing field strength the response time decreases. The standard
Nos\'e--Hoover thermostat Eqs.~(\ref{nh}) is contained as a special
case in equilibrium .

The {\em second variation} goes back to Hoover \cite{Ho89}. It
includes a control of $<p^4>=8T^2$, 
\begin{eqnarray}
\dot{\vec{q}} &=& \vec{p} \nonumber\\
\dot{\vec{p}} &=& \vec{\varepsilon}-\zeta_1\vec{p}
-\zeta_2\frac{p^2}{2T}\vec{p}\nonumber\\ 
\dot{\zeta_1}&=&(\frac{p^2}{2T}-1)\frac{1}{\tau^2}\nonumber\\
\dot{\zeta_2}&=&\frac{p^2}{2T}(\frac{p^2}{2T}-2)\frac{1}{\tau^2}\; .
\label{v1}
\end{eqnarray}
As mentioned in the previous sections this variation can improve
statistical dynamical properties, like ergodicity. 

The {\em third variation} controls $p_x^2$ and $p_y^2$
separately. This is performed by defining two independent reservoirs
for the $x$-- and $y$--direction,
\begin{eqnarray}
\dot{\vec{q}} &=& \vec{p} \nonumber\\
\dot{p_x} &=& \varepsilon_x-\zeta_x p_x \nonumber\\
\dot{p_y} &=& \varepsilon_y-\zeta_y p_y \nonumber\\
\dot{\zeta_x}&=&(\frac{p_x^2}{T}-1)\frac{1}{\tau^2_x}\nonumber\\
\dot{\zeta_y}&=&(\frac{p_y^2}{T}-1)\frac{1}{\tau^2_y}\; .
\label{v2}
\end{eqnarray}
This variation more deeply intervenes in the microscopic dynamics by
forcing the single components $p_x^2$ and $p_y^2$ separately towards
canonical distributions. However, in contrast to the previous
variations a curiosity is hidden in it: At a collision the thermostat
$\zeta_x,\zeta_y$ is uncorrelated to the thermostated variables $p_x$
and $p_y$, because $p_x$ and $p_y$ change at a collision whereas
$\zeta_x,\zeta_y$ remain the same. Therefore the thermostat does not
work efficiently.  But we have not found any reflection of this
curiosity in the macroscopic behavior.

\begin{figure}
\epsfxsize=12cm
\centerline{\epsfbox{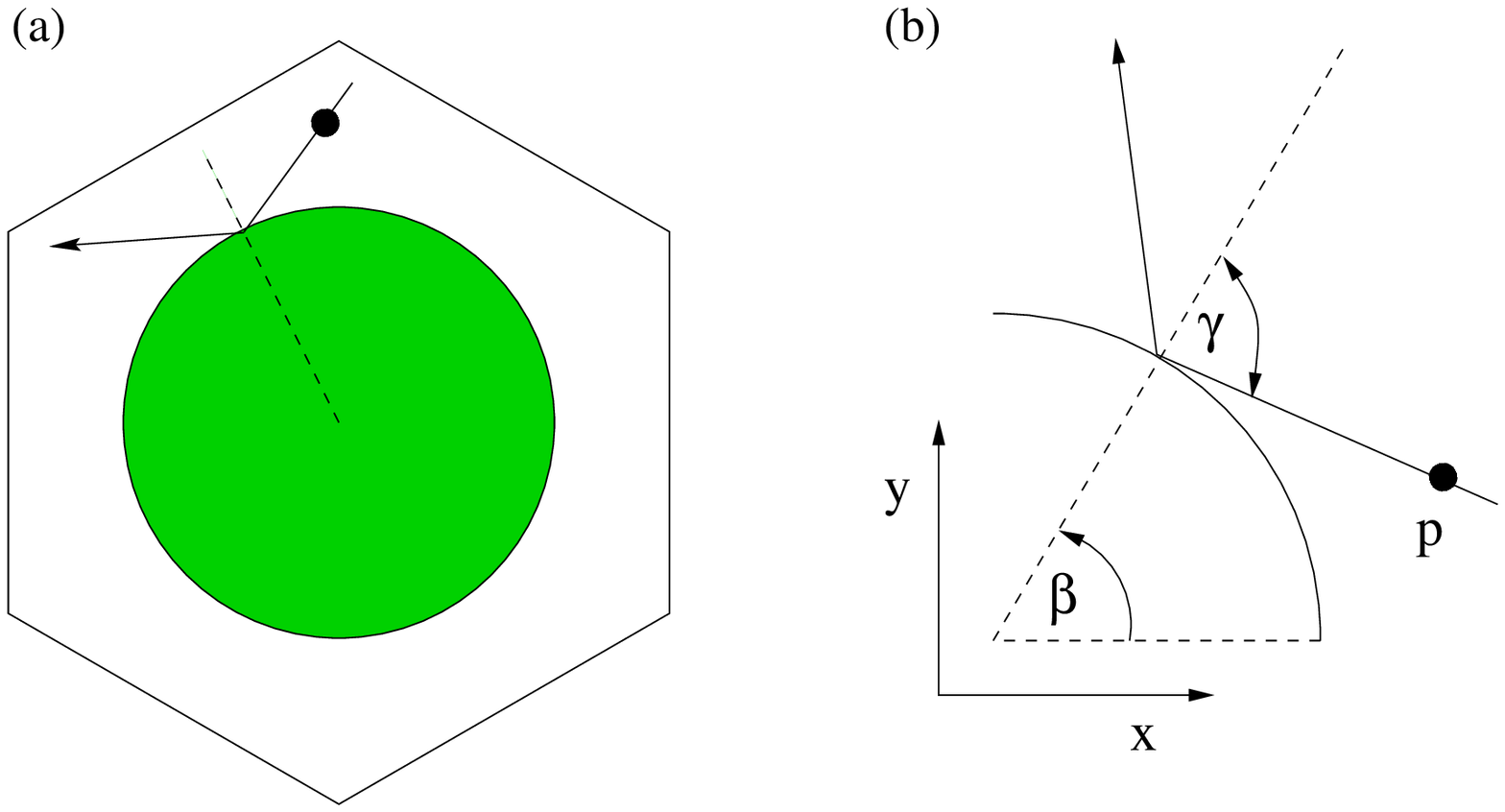}}
\vspace*{0.3cm} 
\caption{(a) Elementary cell of the periodic Lorentz gas on a
triangular lattice. (b) Definition of the relevant variables.}
\label{cell}
\end{figure}

We study the dynamics of these models in one Lorentz gas cell with
periodic boundaries, see Fig.~\ref{cell}(a). As the radius of the disk
we take $r=1$. For the spacing between two neighboring disks we choose
$w\simeq 0.2361$, corresponding to a density equal to $4/5$ of the
maximum packing density of the scatterers. A collisionless free flight
of the particle is avoided for this parameter \cite{MH87}.  The
relevant variables of the dynamical system are defined in
Fig.~\ref{cell}(b): $\beta$ is the angular coordinate of the point at
which the particle elastically collides with the disk, and $\gamma$ is
the angle of incidence at this point.

The equations of motion are integrated by a fourth order Runge Kutta
algorithm with a step size of $dt=0.005$ between two collisions. The
collision of the particle with the disk has been determined with a
precision of $10^{-7}$.  Unless declared otherwise the temperature is
chosen to $T=0.5$.

\section{Equilibrium}
Inserting the initial condition $(p_0,\zeta_0)=(\sqrt{2T},0)$ in
Eqs.~(\ref{nh}) with $\varepsilon=0$ the velocity of the particle
becomes a constant, $p=\sqrt{2T}$, that means the Nos\'e--Hoover
thermostat does not act in the Lorentz gas and the dynamics is
microcanonical. For other initial conditions one observes in computer
simulations that $p^2$ and $\zeta$ oscillate periodically and that the
dynamics of this one particle system is nonergodic.

A stability analysis confirms the numerical results: The
Nos\'e--Hoover thermostated equations of motion Eqs.~(\ref{nh}) can be
reduced for $\varepsilon=0$ to
\begin{eqnarray}
\dot{p^2} &=& -2\zeta p^2 \nonumber\\
\dot{\zeta}&=&(\frac{p^2}{2T}-1)\frac{1}{\tau^2} \; .
\label{nh1}
\end{eqnarray}
Eqs.~(\ref{nh1}) are also valid at the moment of a collision, because
$p^2$ and $\zeta$ are not changed by a collision.  The fixed point of
Eqs.~(\ref{nh1}) is $(p^2,\zeta)=(2T,0)$ with the eigenvalues
$\lambda_{1/2}=\pm\sqrt{-4T/\tau^2}$ and is thus elliptic.

\begin{figure}[htbp]
\epsfxsize=14cm
\centerline{\epsfbox{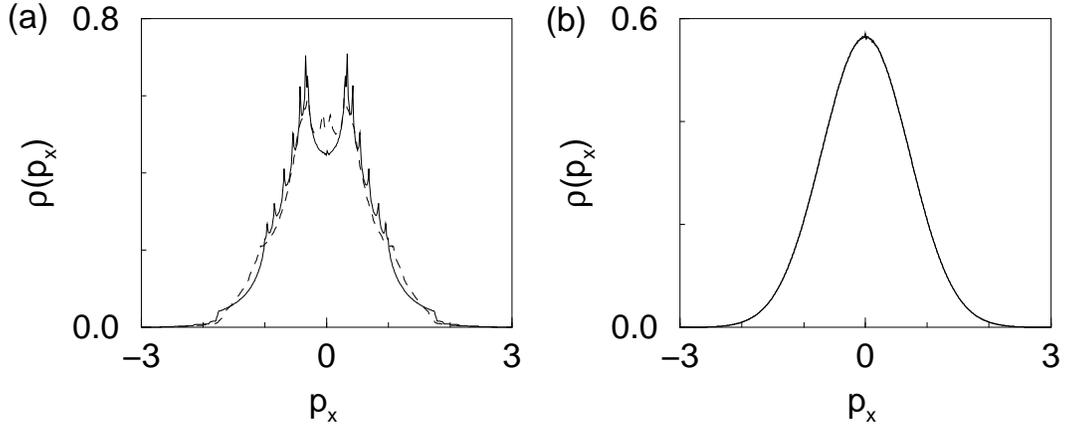}}
\caption{Probability density $\varrho(p_x)$ for $\varepsilon=0$: (a)
additional control of $<p^4>$ with $\tau^2=1$ at two different initial
conditions, (b) control of $<p_x^2>$ and $<p_y^2>$ separately with
$\tau^2_x=0.1$ and $\tau^2_y=1000$}
\label{vxe}
\end{figure}

The additional control of $<p^4>$ destroys the microcanonical
probability density but it is not sufficient to obtain an ``exact''
dynamics\footnote{The dynamics is ``exact'' if every initial density
of nonzero measure converges to the same stationary density
\cite{Mc89}} in the periodic Lorentz gas. Different initial conditions
still lead to a different shape of the probability density
$\varrho(p_x)$, as is shown in Fig.~\ref{vxe}(a).

In contrast the separate control of $p_x^2$ and $p_y^2$ leads to an
exact dynamics in equilibrium corresponding to the canonical
probability density $\varrho(p_x)$, as shown in Fig.~\ref{vxe}(b).

\section{Nonequilibrium}
We now apply an external electric field $\vec{\varepsilon}$ parallel
to the $x$--axis.  The Nos\'e--Hoover thermostat and the related
models then lead to well defined nonequilibrium steady states with
constant average energy of the particle.

\subsection{Probability density $\varrho(p_x)$}
\begin{figure}[htbp]
\epsfxsize=14cm
\centerline{\epsfbox{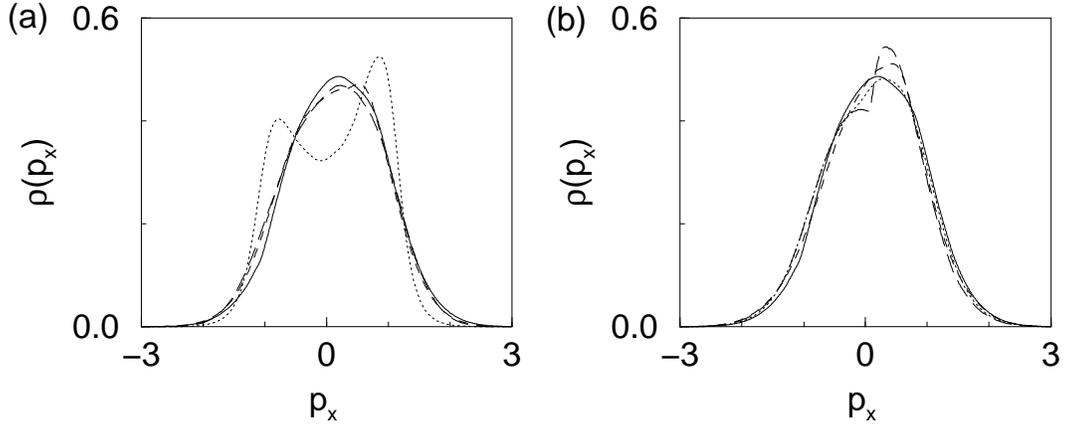}}
\caption{Probability density $\varrho(p_x)$ for $\varepsilon=0.5$: (a)
Nos\'e--Hoover thermostat with $\tau^2=0.01$ (dotted curve),
$\tau^2=1$ (dashed curve), $\tau^2=1000$ (long dashed curve) and
thermostating by deterministic scattering (solid curve), (b)
Nos\'e--Hoover thermostat with field dependent coupling to the
reservoir with $\tau^2=1$ (dotted curve), additional control of
$<p^4>$ with $\tau^2=1$ (dashed curve), control of $<p_x^2>$ and
$<p_y^2>$ separately with $\tau^2_x=0.1$ and $\tau^2_y=1000 $ (long
dashed curve) and thermostating by deterministic scattering (solid
curve).}
\label{vxn}
\end{figure}

The probability density $\varrho(p_x)$ for the Nos\'e--Hoover
thermostat for $\varepsilon=0.5$ is presented in
Fig.~\ref{vxn}(a). For $\tau^2=0.01$ the density shows some remains of
the deformed microcanonical density of the Gaussian thermostat,
whereas for $\tau^2=1$ and $\tau^2=1000$ the density becomes similar
to the density of thermostating by deterministic scattering, which is
related to a canonical distribution \cite{KRN98,RKN99}.

Fig.~\ref{vxn}(b) shows $\varrho(p_x)$ for the three variations of the
Nos\'e--Hoover thermostat. We have chosen here $T=0.60029$ which
corresponds to the temperature in the bulk for thermostating by
deterministic scattering at a parametric temperature of $T=0.5$. The
density of the Nos\'e--Hoover thermostat with field dependent coupling
to the reservoir is very close to the density of thermostating by
deterministic scattering. The density of the Nos\'e--Hoover thermostat
with separate control of $p_x^2$ and $p_y^2$ looks like a
superposition of the densities of the Nos\'e--Hoover thermostat for
small and for large $\tau$.

In all models the mean value of $\varrho(p_x)$ is positive, indicating
a current parallel to the field direction.

\subsection{Conductivity}
\begin{figure}[htbp]
\epsfxsize=14cm
\centerline{\epsfbox{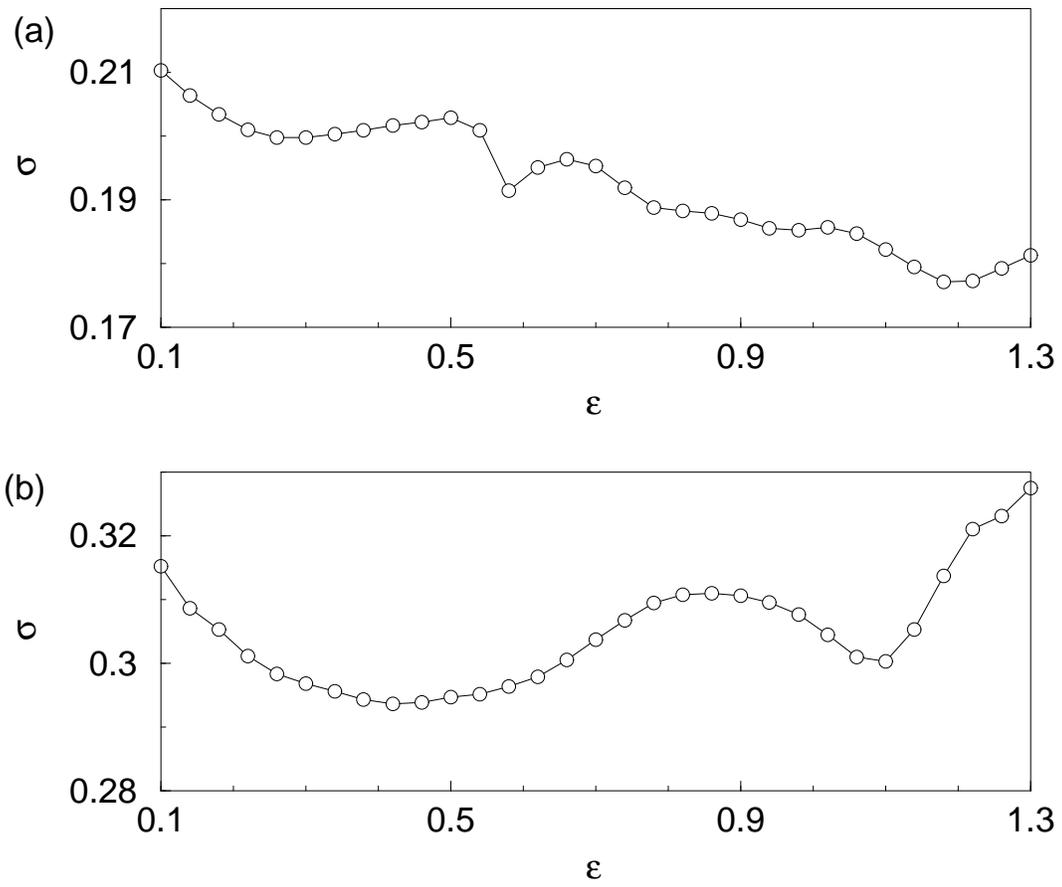}}
\caption{Conductivity $\sigma$ for the Nos\'e--Hoover thermostat, (a)
$\tau^2=0.01$, (b) $\tau^2=1000$. The numerical uncertainty of each
point is less then symbol size.}
\label{cnh}
\end{figure}

The conductivity $\sigma=<p_x>/\varepsilon$ for the Nos\'e--Hoover
thermostat is shown in Fig.~\ref{cnh}. For $\tau^2=0.01$ the curve is
very similar to the conductivity of the Gaussian thermostated Lorentz
gas \cite{LNRM95}. For $\tau^2=1000$ the curve is more stretched along
the $\varepsilon$--axis and globally not decreasing anymore.  In
contrast the conductivity as obtained from thermostating by
deterministic scattering \cite{KRN98,RKN99} is a globally decreasing
function. According to the Einstein relation, in the limit
$\varepsilon\to0$ $\sigma$ should approach the equilibrium diffusion
coefficient $D$ of the periodic Lorentz gas, which for $w=0.2361$ has
the value $D\approx 0.21$ \cite{DeGP95}. This is hard to see for
$\tau^2=1000$, because for $\varepsilon\to 0$ the probability density
changes drastically from a smooth, canonical like density to a
non-smooth density.  It is also difficult to see any linear response
in computer simulations, as has already been discussed for the
Gaussian thermostated Lorentz gas and for thermostating by
deterministic scattering in Ref.~\cite{RKN99}.

\subsection{Attractor}
\begin{figure}[htbp]
\epsfxsize=14cm
\centerline{\epsfbox{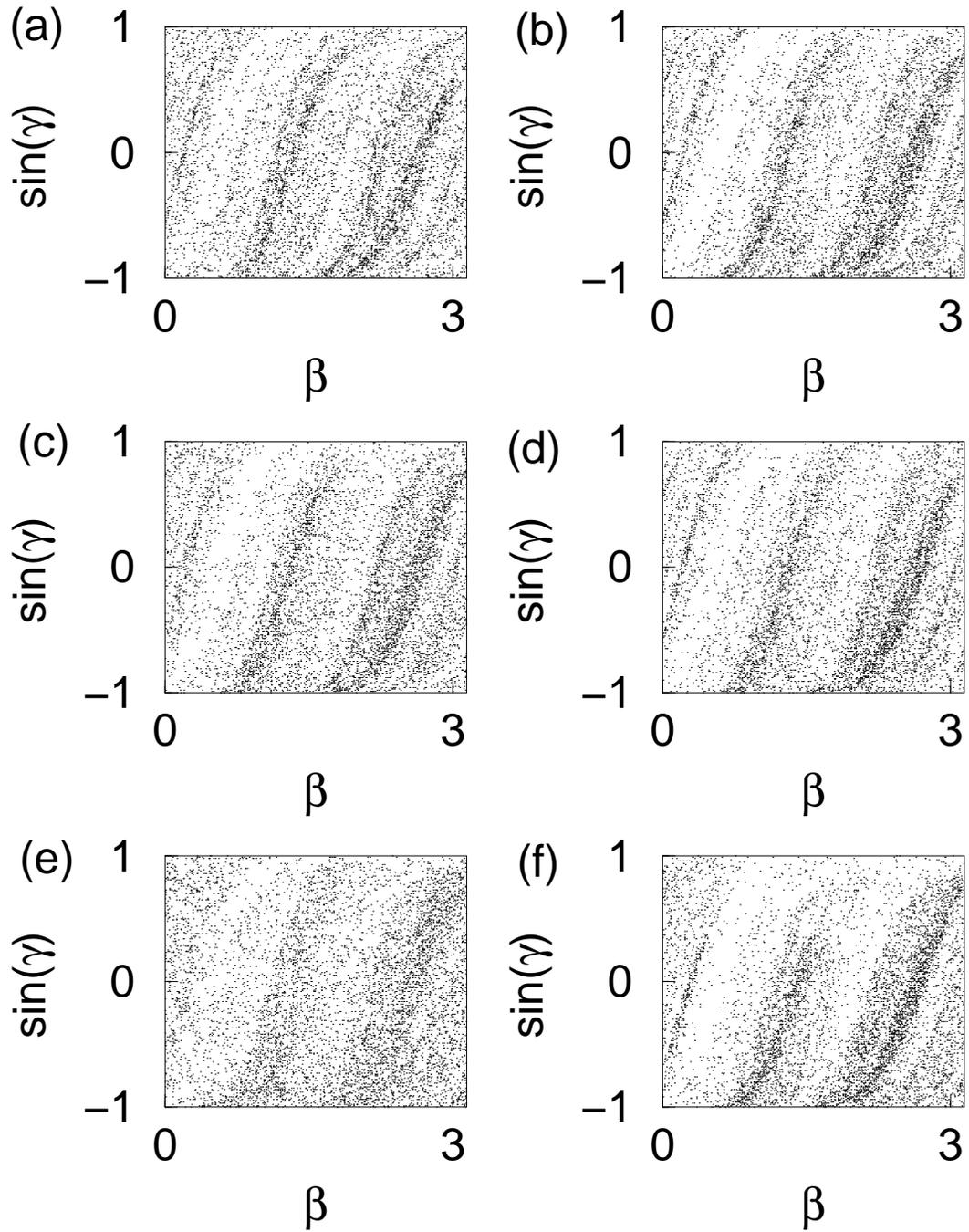}}
\caption{Poincar\'e section of $(\beta,\sin\gamma)$, as defined in
Fig.~\ref{cell} at the moment of the collision, for field strength
$\varepsilon=1$: Nos\'e--Hoover thermostat (a) $\tau^2=0.01$, (b)
$\tau^2=1000$, (c) Nos\'e--Hoover thermostat with field dependent
coupling to the reservoir, $\tau^2=1$ (d) additional control of $p^4$,
$\tau^2=1$, (e) control of $p_x^2$ and $p_y^2$ separately,
$\tau^2_x=0.1$ and $\tau^2_y=1000$, and (f) thermostating by
deterministic scattering.}
\label{pc}
\end{figure}

Fig.~\ref{pc} shows the Poincar\'e section of
$(\beta,\sin(\gamma))$\footnote{Because the symbols for the angles
vary in the literature we mention again that the angle $\beta$ gives
the location of the collision relative to the field direction and
$\gamma$ is the angle of incidence.}  at the moment of the collision
for the Nos\'e--Hoover thermostat, for the three variations and for
thermostating by deterministic scattering. Again, we have chosen
$T=0.60029$ to compare the results with thermostating by deterministic
scattering.  For all models the structure of the attractor is
qualitatively the same as the structure of the fractal attractor
obtained for the Gaussian thermostated Lorentz gas \cite{MH87}.
However, the fine structure varies with the models and with the
response time $\tau$. For the Nos\'e--Hoover thermostat with
$\tau^2=0.01$, see Fig~\ref{pc}(a), the structure is most pronounced,
whereas for the control of $p_x^2$ and $p_y^2$ separately, see
Fig.~\ref{pc}(e), the structure is least visible.

\subsection{Bifurcation diagram}
\begin{figure}[htbp] 
\epsfxsize=18cm
\centerline{\epsfbox{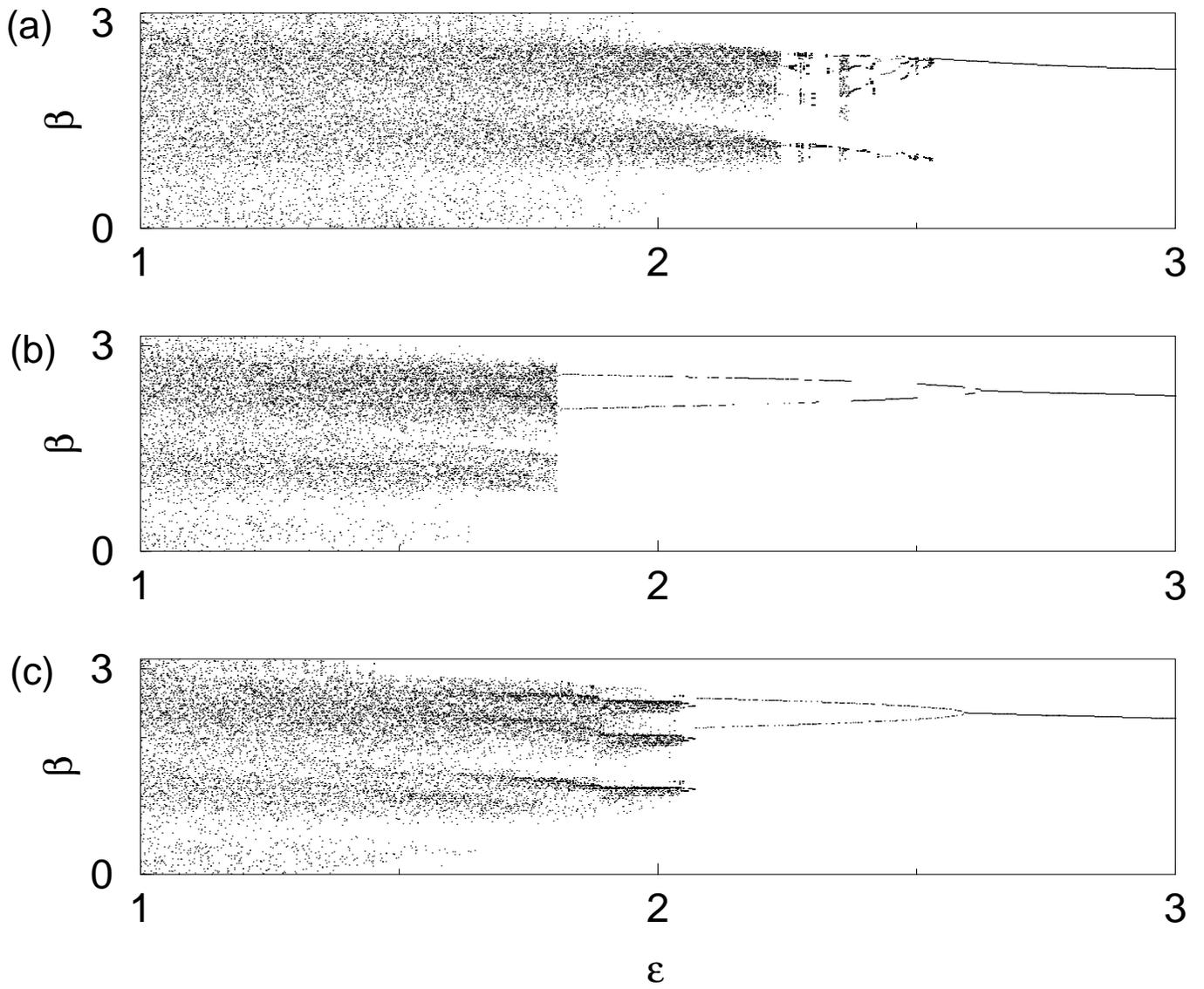}}
\caption{bifurcation diagram for the Nos\'e--Hoover thermostat 
(a) $\tau^2=0.01$, (b) $\tau^2=1$, (c) $\tau^2=1000$}
\label{anh}
\end{figure}

The angle $\beta$ at the moment of the collision is presented as a
function of the field strength for the Nos\'e--Hoover thermostat in
Fig.~\ref{anh}. For all three values of $\tau^2$ the attractor is
phase space filling for small field strengths $\varepsilon<1.3$ and
contracts onto a periodic orbit with increasing field strength.  For
$\tau^2=0.01$ the scenario is similar to the one of the Gaussian
thermostated Lorentz gas \cite{LNRM95}\footnote{In Fig.~10 of
\cite{LNRM95} the angle of flight after a collision is plotted and the
field is parallel to the negative x-axis.}. For $\tau^2=1$ the
scenario looses its richness, but it gets a bit more complicated again
for $\tau^2=1000$. In contrast to the Nos\'e--Hoover thermostat, the
attractor of thermostating by deterministic scattering remains phase
space filling even for large $\varepsilon$ \cite{KRN98,RKN99}.

\begin{figure}[htbp]
\epsfxsize=14cm
\centerline{\epsfbox{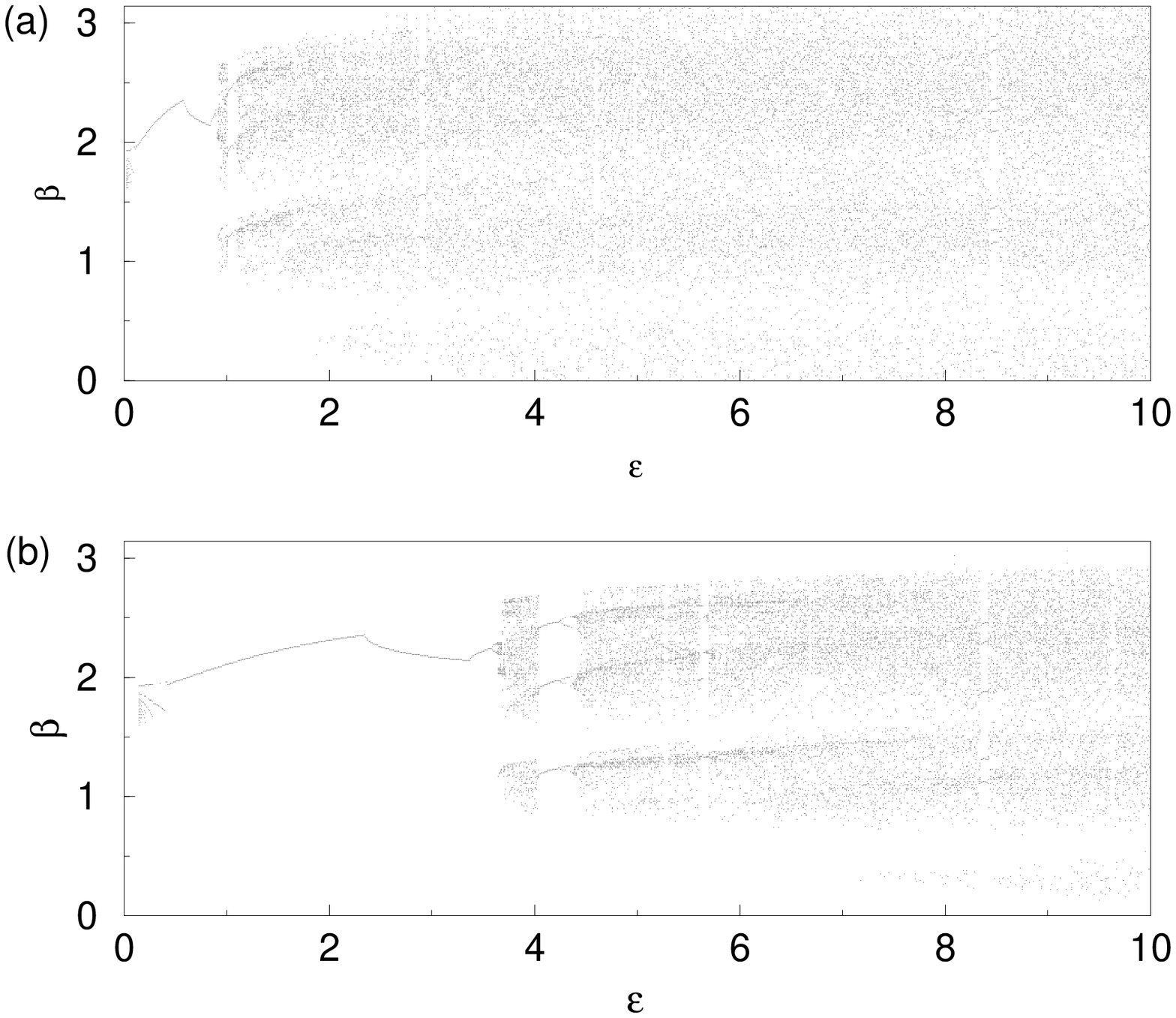}}
\caption{bifurcation diagram for the dissipative limit (a)
$\zeta_c=1.0$, (b) $\zeta_c=2.0$} 
\label{ad}
\end{figure}

The bifurcation diagram in the dissipative limit $\tau\to\infty$ of
Eqs.~(\ref{nh}) with a constant friction coefficient $\zeta_c$ shows
an inverse scenario to Fig.~\ref{anh}, as is presented in
Fig.~\ref{ad}.  For small $\varepsilon$ the trajectory is a so--called
creeping orbit \cite{HoMo92}, then it changes to a periodic orbit, and
for large $\varepsilon$ the attractor gets phase space filling.  By
increasing $\zeta_c$ the strength of the dissipation increases with
the consequence that the onset of chaotic behavior starts at higher
field strengths.  With respect to the numerics we remark that related
to the different shapes of the attractors by varying $\zeta_c$ in the
dissipative limit, the duration of the transient behavior of the
Nos\'e--Hoover thermostat grows drastically for $\tau\to \infty$.

\begin{figure}[htbp]
\epsfxsize=14cm
\centerline{\epsfbox{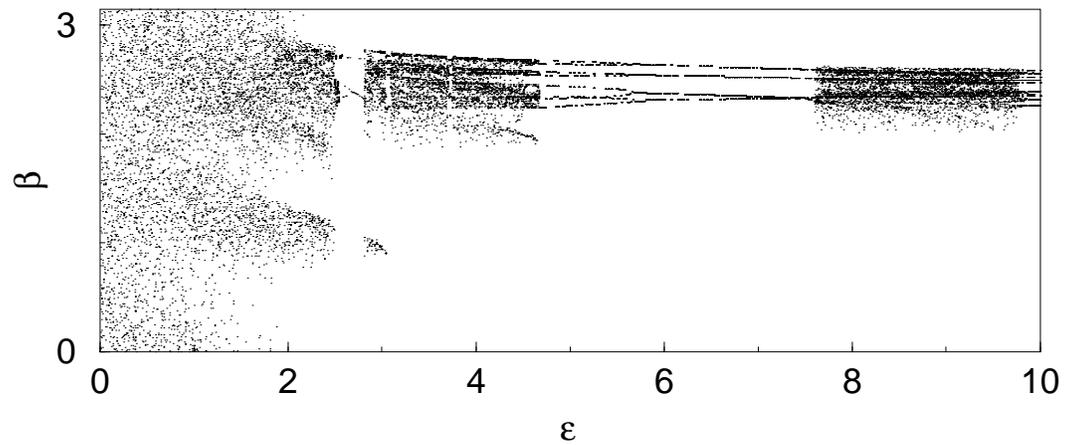}}
\caption{bifurcation diagram for the variation with field dependent
coupling to the reservoir, $\tau^2=1$}
\label{anhe}
\end{figure}

\begin{figure}[htbp]
\epsfxsize=14cm
\centerline{\epsfbox{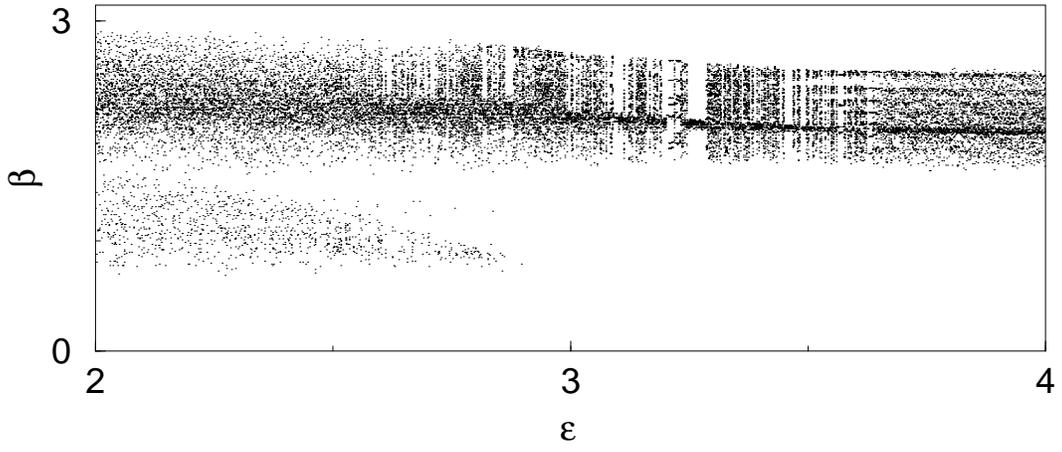}}
\caption{bifurcation diagram for the additional control of $p^4$, $\tau^2=1$}
\label{av1}
\end{figure}

\begin{figure}[htbp]
\epsfxsize=14cm
\centerline{\epsfbox{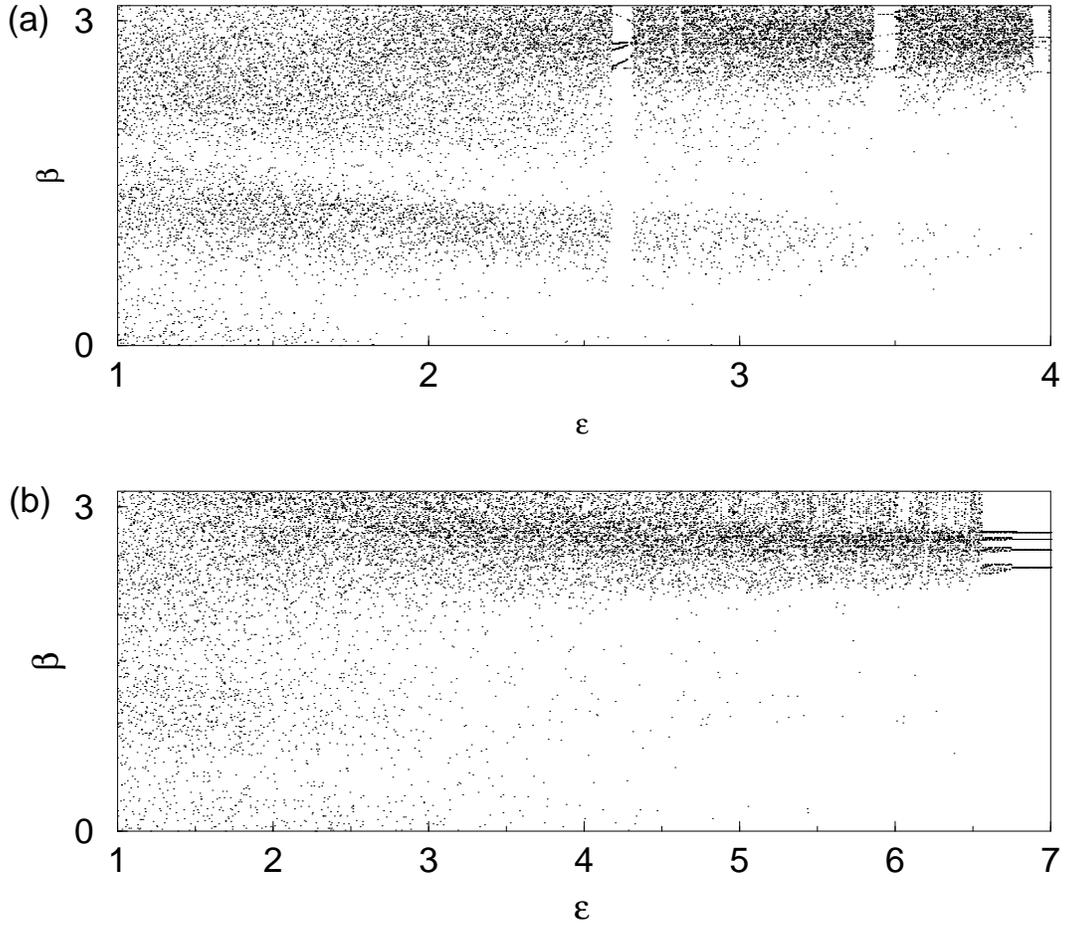}}
\caption{bifurcation diagram for the control of $p_x^2$ and $p_y^2$
separately, (a) $\tau^2_x=\tau^2_y=1$, (b) $\tau^2_x=0.1$,
$\tau^2_y=1000$}
\label{av2}
\end{figure}

Figs.~\ref{anhe}-\ref{av2} show the bifurcation diagrams for the three
variations of the Nos\'e--Hoover thermostat.  In general one observes
that even for high field strengths chaotic regions appear, in contrast
to the Nos\'e--Hoover thermostat.

For the variation with the additional control of $p^4$ the attractor
covers a bounded $\beta$--interval for $\varepsilon\ge 3$. For these
field strengths the trajectory is a creeping orbit.

Fig.~\ref{av2}(b) depicts the attractor for the variation with
separate control of $p_x^2$ and $p_y^2$ under different response times
for the $x$-- and $y$--direction. Since the field acts in the
$x$--direction we have chosen a strong coupling, $\tau^2_x=0.1 $, for
the $x$--direction and a weak coupling, $\tau^2_y=1000$, for the
y--direction.  Up to a field strength $\varepsilon\approx 6.5$ no
periodic window has been found.  The bifurcation diagram for these
parameters most strongly deviates from the Nos\'e--Hoover and Gaussian
thermostated Lorentz gas and is closest to the bifurcation diagram of
thermostating by deterministic scattering. However, in contrast to
thermostating by deterministic scattering the attractor is more
concentrated around $\beta\simeq\pi$.

We detected a numerical problem for the second and the third variation
at several values of $\tau$.
One observes that sporadically after large time intervals there
appears a creeping orbit with a very low velocity of the particle,
which is difficult to handle numerically. Whether this creeping orbit
is the stationary state could probably be clarified by calculating
Lyapunov exponents \cite{DeGP95}.

\subsection{Thermodynamic entropy production and phase space volume
contraction} A characteristic property of the Nos\'e--Hoover
thermostat as well as of the Gaussian isokinetic thermostat is that
the thermodynamic entropy production is equal to the phase space
volume contraction rate \cite{HHP87}.  This equality can as well
easily be verified for the variation with the additional control of
$p^4$ and for the variation with separate control of $p_x^2$ and
$p_y^2$.

On the other hand it does not hold for the Nos\'e--Hoover thermostat
with field dependent coupling to the reservoir. From Eqs.~(\ref{nhe})
one gets for the phase space contraction rate of this model
$-<\rm{div}\dot{\Gamma}>=(2+\varepsilon)<\zeta>$ where
$\dot{\Gamma}=(\dot{\vec{q}},\dot{\vec{p}},\dot{\zeta})$. The precise
relation between thermodynamic entropy production
$\dot{S}_{TD}=\varepsilon<p_x>/T$ and $-<\rm{div}\dot{\Gamma}>$ for
this variation is obtained by calculating the energy balance between
subsystem and reservoir:
\begin{equation}
E=\frac{p^2}{2}+T\tau^2\zeta^2
\end{equation}
is the total energy which in a nonequilibrium steady state should on
average be zero,
\begin{equation}
<\frac{dE}{dt}>=0\;.
\label{en}
\end{equation}
Inserting Eqs.~(\ref{nhe}) with $\varepsilon_y=0$ in Eqs.~(\ref{en})
leads to
\begin{equation}
\frac{\varepsilon_x<p_x>}{T}=\frac{\varepsilon_x<p_x^2\zeta>}{T}+2<\zeta>\;.
\label{ep}
\end{equation} 
Numerical simulations have shown that $p_x^2$ and $\zeta$ are no
independent quantities in nonequilibrium. If $p_x^2$ and $\zeta$ would
be independent and equipartitioning would be fulfilled, i.~e.,
$<p_x^2>=T$, which is only the case in equilibrium, then
Eq.~(\ref{ep}) would lead to an identity between thermodynamic entropy
production and phase space volume contraction.  The results for 
$-<\rm{div}\dot{\Gamma}>$ and for $\dot{S}_{TD}$ as obtained from
computer simulations for this system are presented in Table I at
different $\tau$ and $\varepsilon$. More details of the entropy
production in this variation and in a Gaussian thermostat with the
same property are discussed in Ref.~\cite{KR99}.

\section{conclusions}
We have investigated the Nos\'e--Hoover thermostat and three
variations of it for the periodic Lorentz gas. All models are time
reversible and lead to well defined nonequilibrium steady states with
a constant average kinetic energy of the moving particle.

As a typical characteristic of deterministic and time reversible
thermostating mechanisms it has been confirmed that in nonequilibrium
all these systems contract onto attractors similar to the fractal
attractor of the Gaussian thermostated Lorentz gas.

In equilibrium only the variation of the Nos\'e--Hoover thermostat
with separate control of $p_x^2$ and $p_y^2$ leads to an exact
dynamics with $\varrho(p_x)$ being canonical just like the
corresponding density of thermostating by deterministic scatterin g.

In nonequilibrium the attractor of the Nos\'e--Hoover thermostat
contracts onto a periodic orbit for higher field strength, analogous
to the Gaussian thermostated Lorentz gas. However, the detailed
scenario depends on the value of the response time $\tau$.  Concerning
the probability density in equilibrium, the attractor, and the
conductivity in nonequilibrium, the properties of the standard
Nos\'e--Hoover thermostat in the periodic Lorentz gas are closer to
the properties of the Gaussian thermostat than to the properties of
thermostating by deterministic scattering, although the Nos\'e--Hoover
thermostat and thermostating by deterministic scattering share the
property of keeping the energy of the particle on average constant in
nonequilibrium, and for both thermostats the probability densities for
the velocity components are related to a canonical probability
density.

Concerning the bifurcation diagrams we find that the dynamics of the
three variations of the Nos\'e--Hoover thermostat are in general
``more chaotic''. Even for higher field strengths there exist
pronounced chaotic regions. The separate control of $p_x^2$ and
$p_y^2$ leads to a phase space filling attractor up to high field
strengths in nonequilibrium. The equilibrium properties and the
bifurcation diagram in nonequilibrium of this model are qualitatively
closest to the properties of thermostating by deterministic scattering
in comparison to the other versions.

For the Nos\'e--Hoover thermostat with field dependent coupling to the
reservoir the thermodynamic entropy production is generically not
equal to the phase space volume contraction in contrast to the other
models.

An important question would be to look for common properties of all
deterministic and time reversible thermostats. So far, only the
existence of a fractal attractor in nonequilibrium appears to be
typical. A more detailed investigation of these thermostating
mechanisms should in particular involve a quantitative comparison, as,
e.~g., by means of computing Lyapunov exponents.  An answer to this
question would be helpful for obtaining a general characterization of
nonequilibrium steady states.

{\bf Acknowledgments:} We dedicate this article to G.~Nicolis, a
champion of chaoticity, on occasion of his 60th birthday. K.~R.~and
R.~K.~thank G.~Nicolis for his continuous support and for the
possibility to collaborate with him on problems of thermostating. This
work has been started during the workshop ``Nonequilibrium Statistical
Mechanics'' (Vienna, February 1999). The authors thank the organizers
G.~Gallavotti, H.~Spohn and H.~Posch for the invitation to this
meeting.  K.~R.~thanks the European Commission for a TMR grant under
contract no.\ ERBFMBICT96-1193. R.~K.~acknowledges as well financial
support from the European Commission.

\newpage

\newpage

\newpage
\begin{table}
\caption{Phase space volume contraction rate and thermodynamic
entropy production for the Nos\'e--Hoover thermostat with field
dependent coupling to the reservoir. The numerical error is $\leq
0.001$}
\vspace{0.5cm}
\begin{tabular}{c|cccc}
$\varepsilon$ & $\tau^2=1$ && $\tau^2=1000$ & \\ &
$-<\rm{div}\dot{\Gamma}>$ & $\dot{S}_{TD}$ & $-<\rm{div}\dot{\Gamma}>$
& $\dot{S}_{TD}$ \\
\hline
0.5 & 0.152 & 0.145 & 0.145 & 0.147 \\ 1.0 & 0.547 & 0.561 & 0.567 &
0.592 \\ 1.5 & 1.240 & 1.366 & 1.256 & 1.391
\end{tabular}
\label{t1}
\end{table}
\end{document}